\documentclass[12pt]{article}
\usepackage[T1]{fontenc}
\usepackage[latin9]{inputenc}
\usepackage[a4paper,text={16cm,25cm},centering]{geometry}
\usepackage{graphicx}
\usepackage{amsmath,amssymb}
\usepackage[numbers]{natbib}

\begin{document}

\title{On the turbulent sources of the solar dynamo}
\author{}
\date{}
\maketitle

\vspace{-2cm}

\begin{center}
V. V. Pipin\\
{\small Institute for Solar-Terrestrial Physics \\ Siberian Division of the Russian Academy of Sciences\\ 664003 Irkutsk, RUSSIA\\
 e-mail address: pip@iszf.irk.ru}
\\[3mm]
N. Seehafer\\
{\small Institut f\"ur Physik und Astronomie,\\ Universit\"at Potsdam,\\ Karl-Liebknecht-Str. 24/25\\
 14476 Potsdam-Golm, GERMANY\\
e-mail address: seehafer@uni-potsdam.de}
\end{center}

\vspace{0pt}

\begin{abstract}

We revisit the possible turbulent sources of the solar dynamo. Studying axisymmetric mean-field
dynamo models, we find that the large-scale poloidal magnetic field could be generated not only
by the  famous $\alpha$ effect, but also by the $\mathbf{\Omega}\times \mathbf{J}$
and shear-current effects.
The inclusion of these additional turbulent sources alleviates several
of the known problems of solar mean-field dynamo models.
\end{abstract}

\section{Introduction}

\begin{figure}
\begin{centering}
\includegraphics[height=6.6cm,angle=-90]{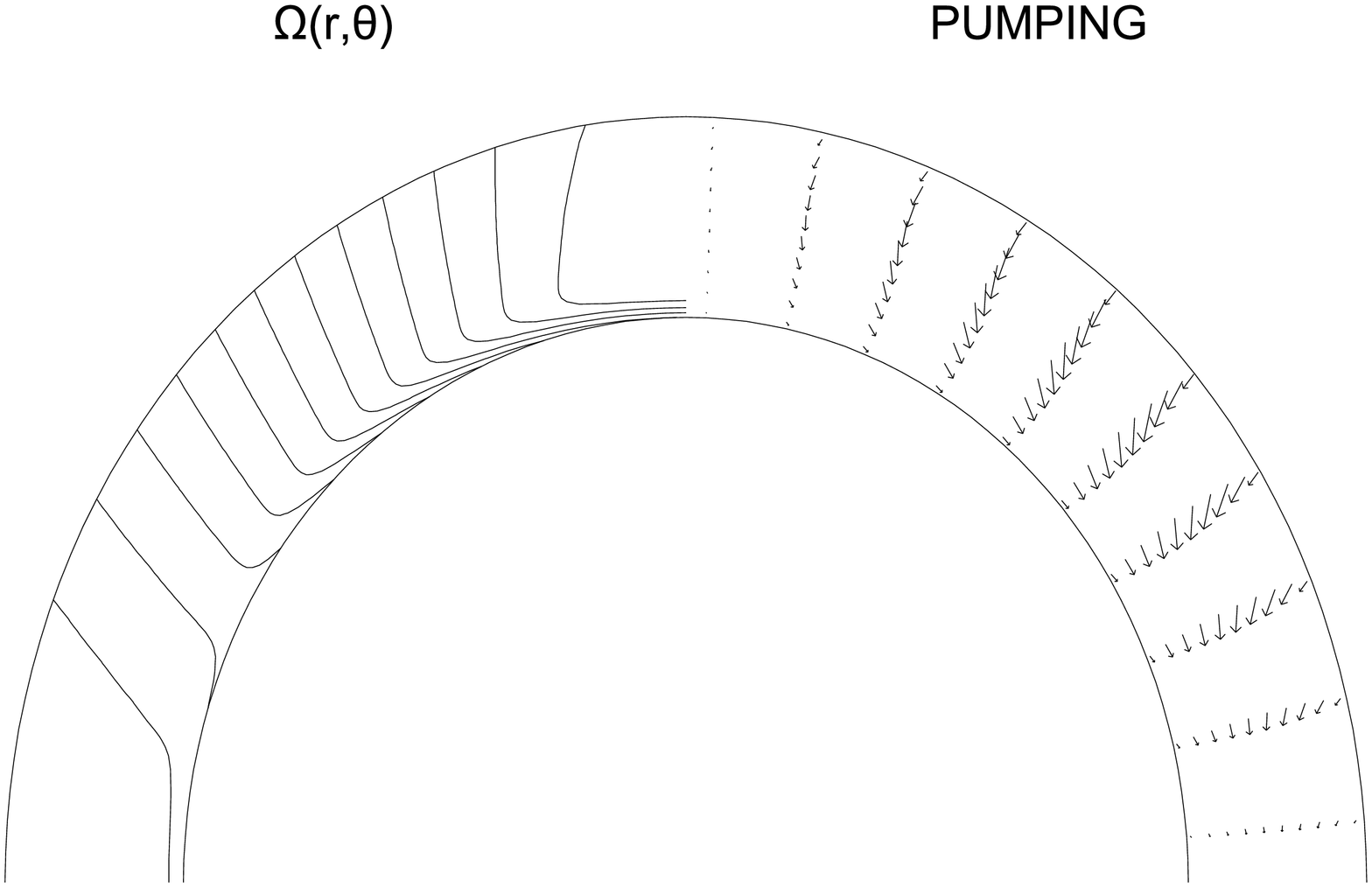}
\includegraphics[height=6.6cm,angle=-90]{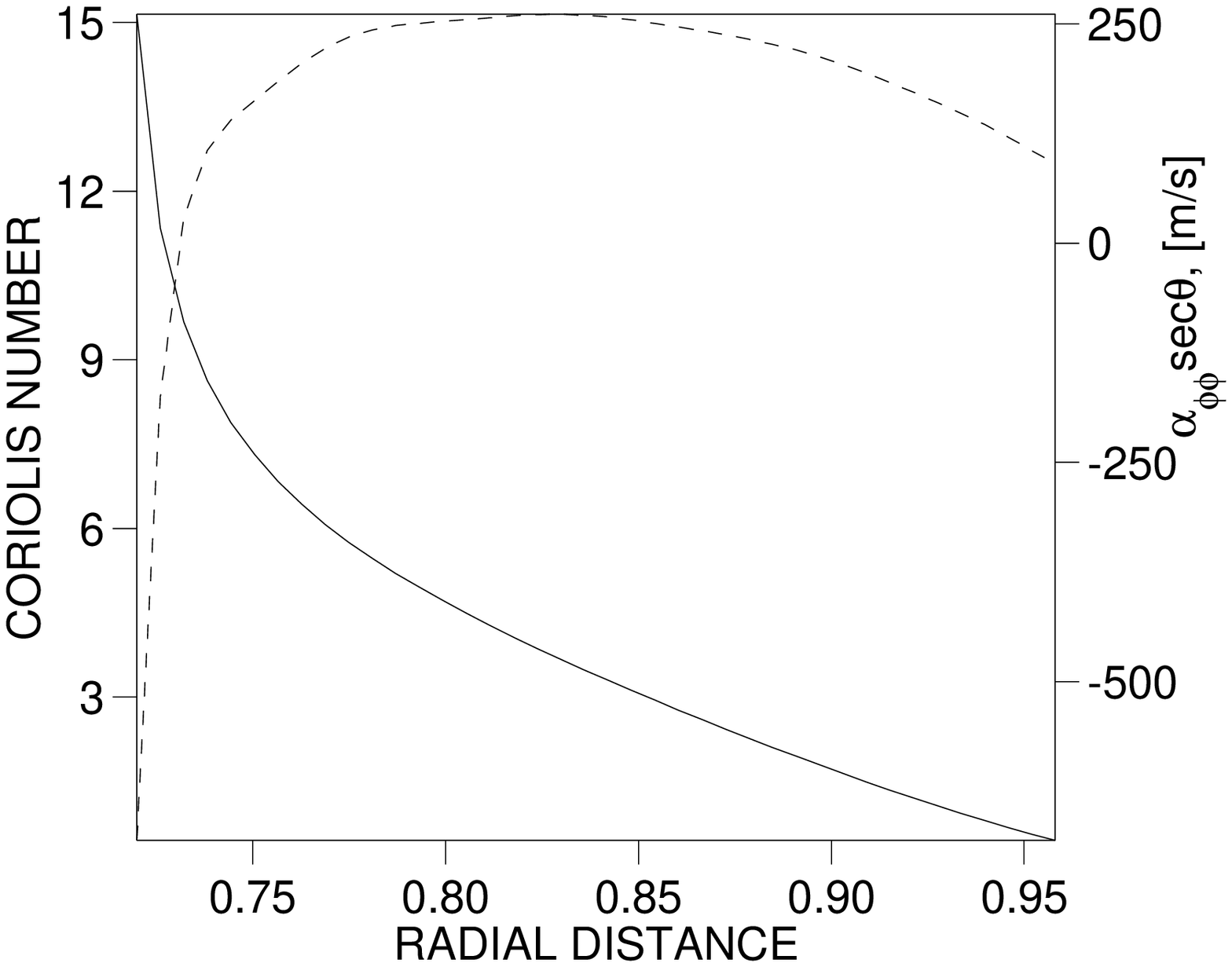}
\caption{ \small\label{fig:fig0}Basic model quantities. Left panel: Contours of the rotation rate
in the solar convection zone (left) and
geometry of the pumping velocity of the toroidal LSMF (right); the maximum amplitude of the
pumping velocity
is about $1\, \mathrm{m}/\mathrm{s}$. Right panel: Radial profiles of the Coriolis
number $\Omega^{*}$ (solid line)
and  $\alpha_{\phi\phi}\sec\theta$ (dashed line).
$\alpha_{\phi\phi}$ changes sign near the bottom
of convection zone. Similar radial dependences are found for the other
components of the $\hat{\alpha}$ tensor.}
\par
\end{centering}
\end{figure}

Most  solar dynamo models use the scenario proposed
by Parker \cite{par55}. In this scenario, the solar magnetic field
is produced by an interplay between differential rotation and the
collective action of  cyclonic turbulent convection flows. The latter is widely
known as the $\alpha$ effect \cite{krarad80}, which is believed
to be responsible for the generation
of the poloidal component of the large-scale magnetic field (LSMF) of
the Sun. There are, however, a number of problems with the $\alpha$ effect \cite{brasub05}, and recent developments of
mean-field magnetohydrodynamics have stimulated a quest for additional or alternative turbulent
dynamo mechanisms. The influence of turbulence on the evolution of
the LSMF is expressed by the mean electromotive
force (MEMF) $\boldsymbol{\mathcal{E}}=\left\langle \mathbf{u}\times\mathbf{b}\right\rangle$, where
$\mathbf{u}$ and $\mathbf{b}$ are the fluctuating parts of the velocity and magnetic
field (angular brackets denote ensemble averages). It can be written in compact form as
\begin{equation}
\boldsymbol{\mathcal{E}}=\left(\hat{\alpha}+\hat{\gamma}\right)\circ\mathbf{B}-\left(\hat{\eta}+\hat{\Omega}+\hat{W}+ \dots \right)\circ\left(\nabla\times\mathbf{B}\right).
\label{eq:emf}\end{equation}
Here $\mathbf{B}$ is the mean, large-scale magnetic field. The tensors
$\hat{\alpha}$, $\hat{\gamma}$ and $\hat{\eta}$ 
describe
the $\alpha$ effect, turbulent pumping and turbulent diffusion, $\hat{\Omega}$
the $\mathbf{\Omega}\times\mathbf{J}$ effect (or $\delta^{(\Omega)}$ effect)
 and $\hat{W}$ the shear-current effect (or $\delta^{(W)}$ effect). Discussions of the
 different contributions to $\boldsymbol{\cal{E}}$ may, for instance, be found in
\cite{radste06,rogkle04}.
In this paper, we use  calculations of them as described in \cite{pip08}, and construct a set of axisymmetric
kinematic dynamo models in a convective spherical shell. These models are
obtained as solutions of the mean-field dynamo equation
\begin{equation}
\frac{\partial\mathbf{B}}{\partial
  t}=\mathbf{\nabla}\times\left(\mathbf{V}\times\mathbf{B}
+\boldsymbol{\mathcal{E}}\right),\label{eq:dyn}
\end{equation}
where $\mathbf{V}=r\sin\theta\,\Omega\left(r,\theta\right)$, with $\Omega(r,\theta)$
modeling the differential angular rotation rate
of the Sun ($r$ and $\theta$ denote radius and colatitude); see Fig.~\ref{fig:fig0}, left-hand side in left panel.
 The LSMF is modeled in the form
$\mathbf{B}=\mathbf{e_{\phi}}B\left(r,\theta\right)+\mathbf{\nabla}\times\left({\displaystyle \frac{A\left(r,\theta\right)\mathbf{e_{\phi}}}{r\sin\theta}}\right)$
($\mathbf{e}_{\phi}$ denotes the
azimuthal unit vector), and the MEMF is calculated
according to Eq.~(\ref{eq:emf}). Details of the procedure are given in \cite{pipsee08}.
The radial profiles of characteristic
quantities of the turbulence, such as the rms convective velocity $u_{c}$, the
correlation length and time $\ell_{c}$ and $\tau_{c}$, and the density stratification
parameter $G=\nabla\log\rho$ %, $U=2\nabla\log u_{c}$]
($\rho$ is the mass density), are computed
on the basis of a standard model of the solar interior \cite{sti02}.
Some of the generation effects depend on the rms intensity 
of a fluctuating background magnetic field $\mathbf{b}^{(0)}$, generated
by a small-scale dynamo.
We assume energy equipartition between $\mathbf{b}^{(0)}$ and the turbulent velocity field, i.e. $\sqrt{\left\langle{\mathbf{b}^{(0)}}^{2}\right\rangle}/\left(u_{c}\sqrt{4\pi\rho}\right)=1$.
The relative strengths of the turbulence effects are controlled via parameters
$C_{\alpha}$ ($\alpha$ effect), $C_{\omega}$ ($\mathbf{\Omega}\times \mathbf{J}$ effect) and $C_{W}$ (shear-current effect); in addition, turbulent pumping is always present, depending merely
on the turbulence level (with which all other contributions are also scaled).
 The integration
domain is radially bounded by $r=0.71R_{\odot}$ and $r=0.96R_{\odot}$,
where the boundary conditions are
${\displaystyle \frac{\partial rB}{\partial r}=0,A=0}$ at the bottom boundary,
and vacuum conditions at the top boundary. Four basic
model quantities, namely, the differential rotation,
the turbulent pumping velocity of the toroidal component of the LSMF and
the radial profiles of the Coriolis number $\Omega^{*}=2\Omega_{0}\tau_{c}$ ($\Omega_{0}=2.86\cdot 10^{-6}s^{-1}$ is the surface rotation rate)
and  of $\alpha_{\phi\phi}\sec\theta$ ($\alpha_{\phi\phi}$ measures the strength of the azimuthal $\alpha$ effect) are shown in Fig.~\ref{fig:fig0}.

\section{Results}

\begin{figure}
\begin{centering}
\includegraphics[width=6cm,angle=-90]{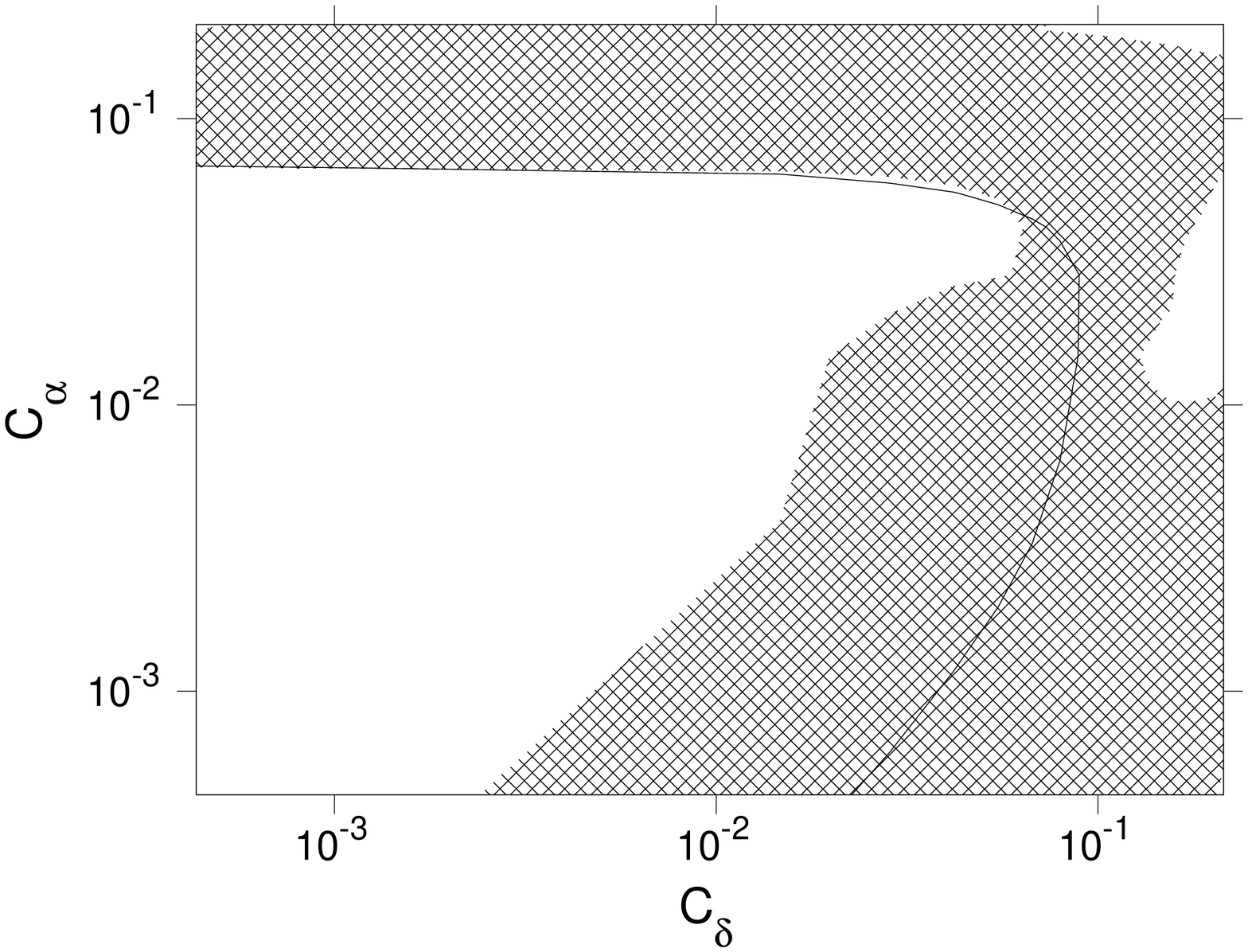}~\includegraphics[width=6cm,angle=-90]{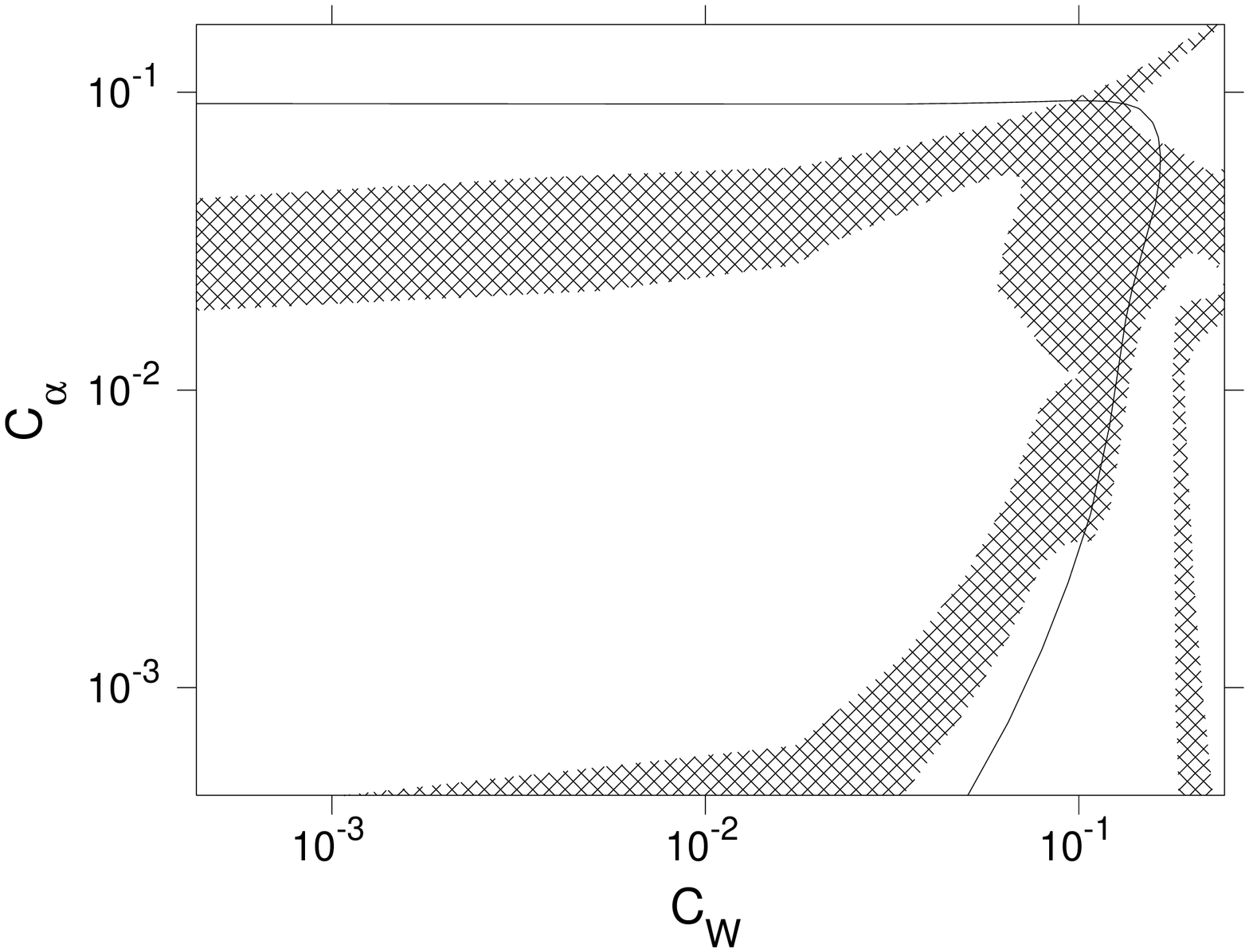}
\par\end{centering}
\caption{ \small\label{fig:stab} Stability diagrams for the
$\alpha^{2}\delta^{(\Omega)}\Omega$ (left panel) and $\alpha^{2}\delta^{(W)}\Omega$ (right panel) dynamos. The stable regions (no dynamo) lie below and to the left of the solid
lines. Shading indicates dominance of dipole modes.
}
\end{figure}

\begin{figure}
\begin{centering}
\includegraphics[angle=-90,width=0.8\textwidth]{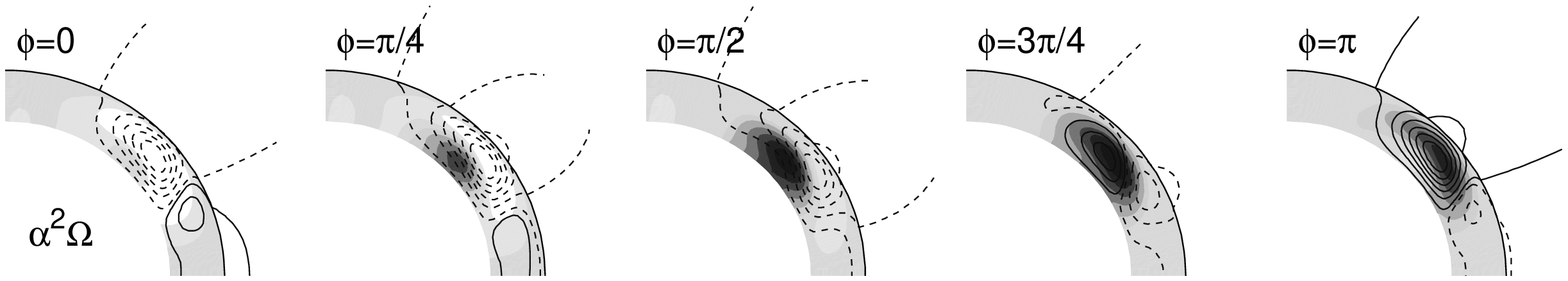}
\par\end{centering}
\begin{centering}
\includegraphics[angle=-90,width=0.8\textwidth]{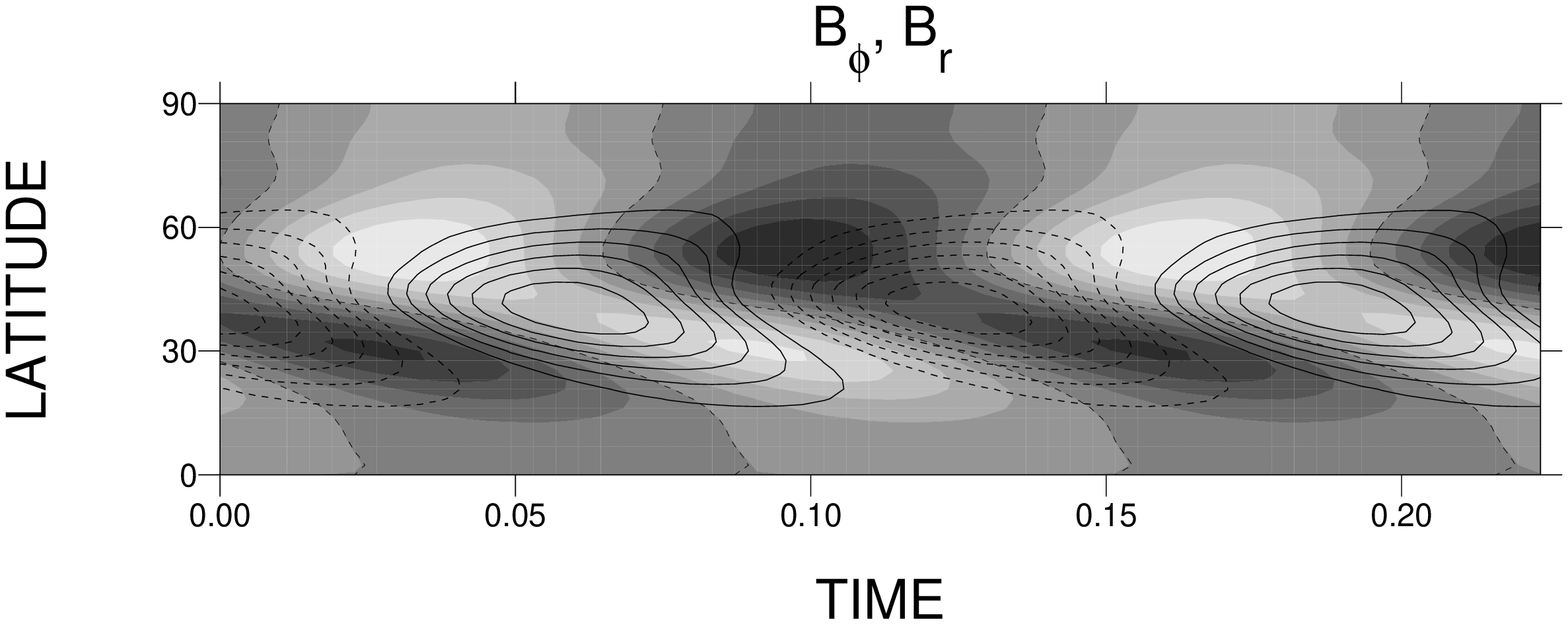}
\par\end{centering}
\begin{centering}
\caption{ \small \label{fig1} $\alpha^{2}\Omega$ model on the basis of the first unstable dipole
eigenmode. Top: Snapshots of the strength of the toroidal LSMF (greyscale
plot) and field lines of the poloidal LSMF over
an half cycle. Bottom: Butterfly diagram in the form of contour lines of the toroidal LSMF (integrated
over radius) with overlaid greyscale plot for the radial
LSMF at the top boundary. Time is measured in units of the turbulent diffusion time.}
\par\end{centering}
\end{figure}

We now discuss examples of dynamos obtained by combining
the $\alpha$ effect with the $\mathbf{\Omega}\times\mathbf{J}$  and
shear-current effects. It was found earlier \cite{sti76} that
for solar conditions and without meridional circulation, $\delta^{(\Omega)}\Omega$
dynamos ($\mathbf{\Omega}\times\mathbf{J}$ effect plus differential rotation) have
steady non-oscillatory dynamo eigenmodes. We found that the same applies to $\delta^{(W)}\Omega$ dynamos
(shear-current effect plus differential rotation) and to combinations of
$\delta^{(\Omega)}\Omega$ and  $\delta^{(W)}\Omega$ dynamos.
This means that the $\alpha$ effect is needed
in solar dynamo models, at least in such without meridional
circulation.
Fig.~\ref{fig:stab} 
gives stability diagrams
for $\alpha^{2}\delta^{(\Omega)}\Omega$ and $\alpha^{2}\delta^{(W)}\Omega$ dynamos,
which include the usual $\alpha^{2}\Omega$ dynamo as a limiting case. As is seen in
Fig.~\ref{fig:stab}, in this latter case the dipole modes are not the most unstable modes.
Fig.~\ref{fig1}
shows snapshots
of the evolution of the LSMF in the $\alpha^{2}\Omega$ dynamo
and the associated butterfly diagram, obtained
from the first unstable dipole eigenmode. We assume that the sunspot
activity is produced by the toroidal LSMF in the whole convection
zone and have integrated the toroidal field over
radius. The obtained dynamo has some similarity with the observed evolution
of the large-scale magnetic activity of the Sun, e.g., the ``correct''
phase relation between the toroidal and poloidal components
--- the polar reversal of the radial component of the LSMF takes place when the
amplitude  of the toroidal component passes its maximum, and
the toroidal fields drift towards the equator in the course of the activity cycle.
There are a number of problems, however. Two of them are: (i) The period of the activity cycle is
$T\lesssim 1\,\mathrm{yr}$ (corresponding to about 0.1 turbulent diffusion times), which is
much shorter than the observed period of 22\,yr.
(ii) The activity maxima
are found at mid latitudes, rather than, as observed, at low latitudes.

\begin{figure}
\begin{centering}
\includegraphics[angle=-90,width=0.8\textwidth]{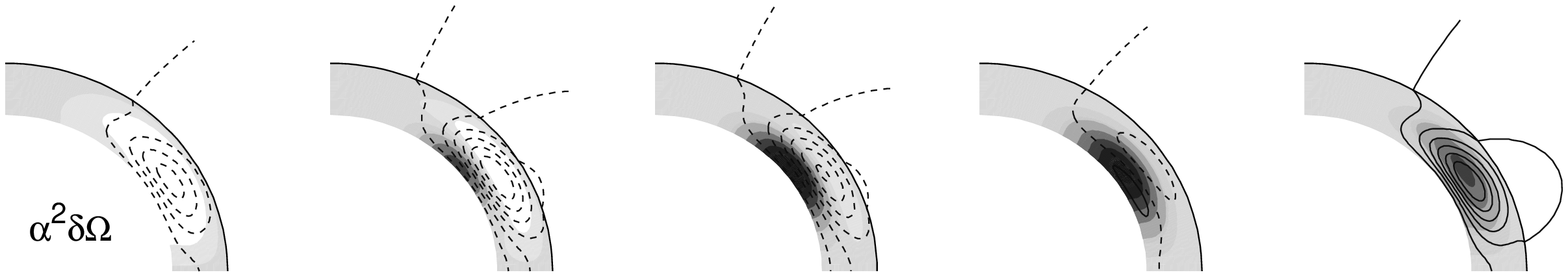}
\par\end{centering}
\begin{centering}
\includegraphics[angle=-90,width=0.8\textwidth]{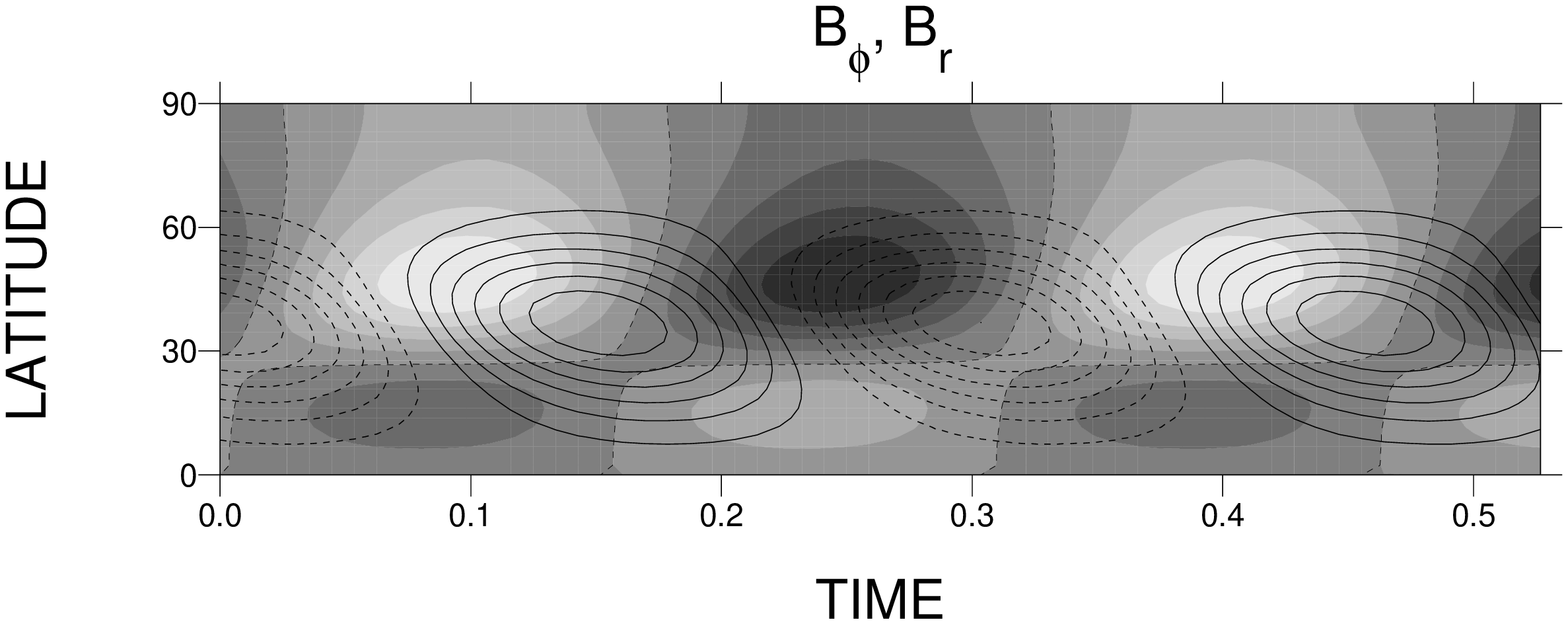}
\par\end{centering}
\begin{centering}
\caption{\label{fig2}  \small As Figure \ref{fig1}, but for the $\alpha^{2}\delta^{(\Omega)}\Omega$ model with $C_{\omega}=0.1$, $C_{\alpha}=0.05$. }
\par\end{centering}
\end{figure}

The inclusion of the two additional turbulent generation mechanisms of the poloidal LSMF
reduces the frequency of the dynamo wave. Fig.~\ref{fig2}
demonstrates this for
the example of an $\alpha^{2}\delta^{(\Omega)}\Omega$ dynamo. Oscillatory modes
are excited if $C_{\alpha}$ is not very small compared to $C_{\omega}$.
In the range
$1<C_{\omega}/C_{\alpha}<3$ the match with the solar observations is best. As can be seen
in Fig.~\ref{fig:stab},
the dipole modes are the primary dynamo modes in this parameter interval.
While the wings of the obtained dynamo waves are too wide, other qualitative
properties, such as the drift directions of the toroidal and poloidal components
of the LSMF and their phase relation, are reproduced correctly. The period of the dynamo in this parameter
range is shorter than but comparable with the turbulent diffusion time.

\begin{figure}
\begin{centering}
\includegraphics[angle=-90,width=0.8\textwidth]{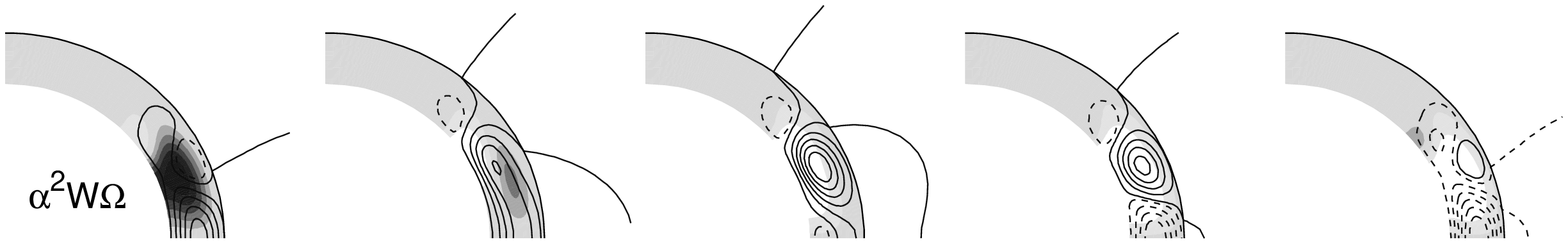}
\par\end{centering}
\begin{centering}
\includegraphics[angle=-90,width=0.8\textwidth]{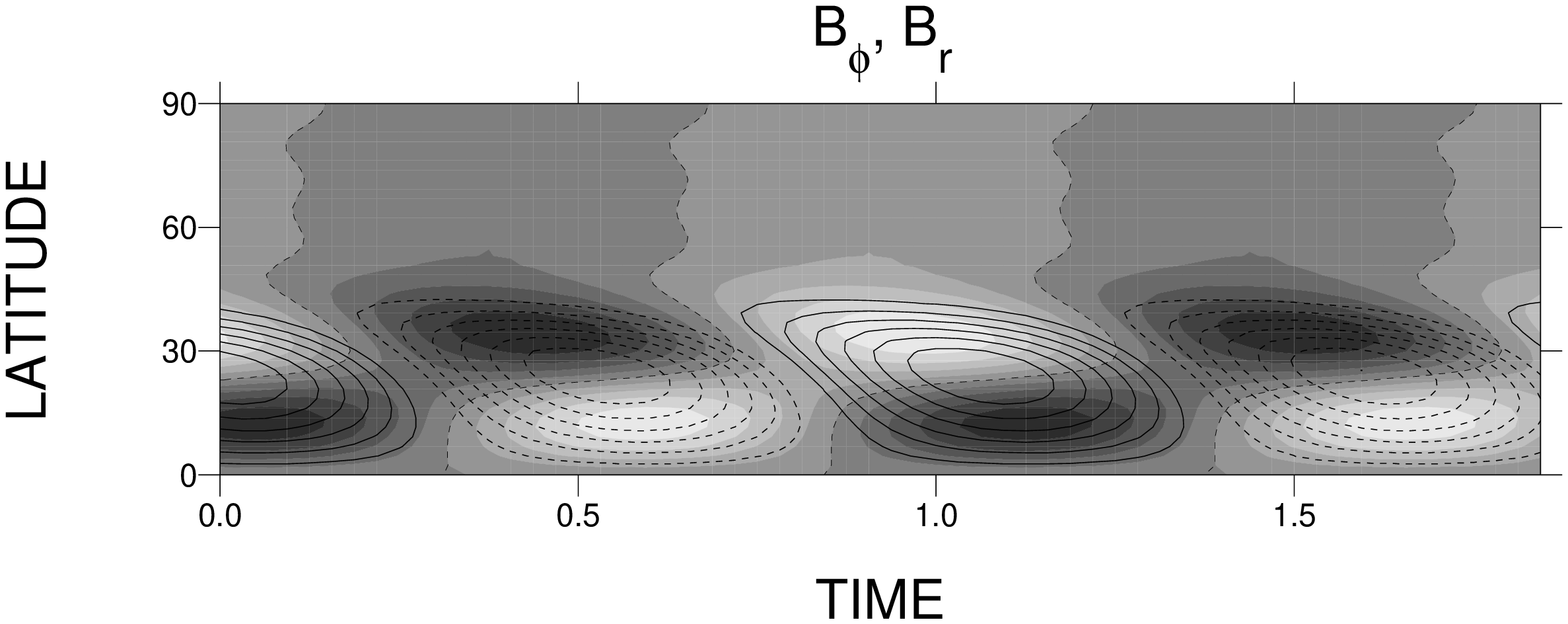}
\par\end{centering}
\begin{centering}
\caption{\label{fig3} \small As Figure \ref{fig1}, but for the $\alpha^{2}\delta^{(W)}\Omega$ model
with $C_{W}=0.1$, $C_{\alpha}=0.01$. }
\par\end{centering}
\end{figure}

Our last example is a dynamo where the poloidal LSMF
is built up with the help of the shear-current effect.
Fig.~\ref{fig3} 
shows the time evolution of the LSMF
and the associated butterfly diagram for an $\alpha^{2}\delta^{(W)}\Omega$
dynamo model. Some properties of the model agree well
with the observations. For example, the period
of the dynamo waves is comparable with the turbulent diffusion time.
 Also, the maximum amplitudes of the toroidal component of the LSMF are found below
$30^{\circ}$ latitude, and this field component drifts towards the equator during the cycle.
A very unwelcome issue of the model, however, is that the maxima of the radial component
of the LSMF are also concentrated near the equator.

\section{Conclusions}

The inclusion of the $\mathbf{\Omega}\times \mathbf{J}$ and shear-current effects
in addition to the $\alpha$ effect and differential rotation in axisymmetric kinematic
dynamo models alleviates some
of the known problems of current mean-field solar dynamo models. For instance,
the simulated period of the dynamo is increased and comes, thus, closer to the observed activity
period.
Furthermore,
the large-scale toroidal field is concentrated towards the equator, thus bringing the models in better
agreement with the observation of two activity belts relatively close to the equator.
Improved diagnostic tools based on both observation and theory are needed to clarify the roles
of the different turbulence effects, as well as that of meridional circulation, which was not included
in the present study.

\bibliographystyle{plainnat}
%\bibliography{/home/seehafer/TEX.DIR/LIB.DIR/genbib}
%\bibliography{/usr/users3/seehafer/TEX.DIR/LIB.DIR/genbib}

\end{document}